# HIERARCHY BASED MICROWORLD SCALES' CLASSIFICATION AND MICROWORLD PHYSICS

A. Ya. Temkin

Department of Interdisciplinary Studies

Faculty of Engineering

Tel-Aviv University

Ramat-Aviv

Tel-Aviv 69978

Israel

E-mail: temkin@eng.tau.ac.il

Fax: +972-3-6410189

Tel: +972-3-6408176

June 30, 2005

## *ABSTRACT*

Scales of the microworld are defined on the grounds of the hierarchy that can be set up within a well ordered finite or infinite countable set of well ordered sets. Indirect measurements in pure mathematical as well as physical meaning are considered as the mean to obtain information on the occurring within each scale. The general concept of the physical laws within a certain scale is defined in the framework of the set theory. The hypothesis is proposed that quarks are not within the same scale as other elementary particles are, but within the following ("smaller", exactly of lower hierarchy) scale. Maybe, in particular, this is the cause of difficulties of the free quark detection? Some other consequences of the abovementioned hypothesis are discussed.

Limitations of the possibility to obtain information on the going on within sets of low hierarchy (in particular, "small" scales of the microworld, in physics) and to transfer it toward the set of the highest hierarchy (in particular, to the macroscopic observer in physics) are found as a consequences of the multi-step character of measurements.

# 1. INTRODUCTION

The contemporary high-energy physics that uses particles of higher and higher energies penetrates to smaller and smaller scales of the microworld. In view of this it would be desirable to define the general concept of **the microworld scale** and to find how the information on the occurring within each scale can be forwarded to the **macroscopic observer** possessing the exclusive ability to obtain, to process and to interpret it. It is necessary also to elucidate whether the total number of microworld scales is *principally* limited or not, *i. e.*, whether this penetration to the depth of the microworld by the high-energy physics cannot/can be continued up to its infinitesimally small scales.

Classical and quantum physics correspond to macroscopic and microscopic scales of the world. The second of them, quantum physics' scale, corresponds to space region of characteristic linear dimension $\stackrel{<}{\sim} 10^{-8} cm$, *i. e.*, atomic dimension or less. Whether "less" means "up to infinitesimally small linear dimension", or there exists its lowest limit $l_{\min,1}$ such that from linear dimension $\stackrel{<}{\sim} l_{\min,1}$ begins the new scale with its own physical laws that stretches up to $l_{\min,2}$, and so on? It is, at least, not impossible. To consider this and other problems of the microworld scaling it is first of all to define the concept "scale". The difficulty is that we cannot be sure that the space-time continuum and, therefore, dimension exists always. By this reason it would be desirable to define the concept *scale* without use the notions of upper and lowest limits of its dimension.

<u>In the present work the starting point is the hierarchy that is set up among well ordered sets forming a well ordered set.</u> The microworld scales are considered as a particular case of this set theoretical consideration. They defined and classified with respect to their hierarchy, which is established, in this case, *on the grounds of properties and characters of physical phenomena, and of the information on them*. This approach to the microworld scales' classification does not demand the use of the notion <u>dimension</u> of each scale (for example, by assignment scale's upper and lowest limits) for the scale definition. Thus, it delivers us from the necessity to use space and time (or space-time) as the area for physical event addresses' representation <u>throughout all scales</u>, as it is being done in classical and quantum mechanics, *i. e.*, within the macroscopic and atomic scales.

**Note**:
1)   In the present paper we, for brevity, call <u>events</u> not only events themselves (*e. g.*, collisions), but also objects (*e. g.*, elementary particles).
2)    When we consider the classical and quantum scales, and only in these cases, we call addresses of physical events their

co-ordinates in space-time, spin, isotopic spin, parity *etc.*.

The general representation of event addresses is introduced.

...................................................

In Sec. 2 the concept of *information types* is defined and used to introduce the concept of set types for sets able to treat the information. It is proposed to define *type* of such a set according that what functions of the information treatment it executes. However, other possibilities of type definition are considered preliminary (detailed consideration is in Secs. 3 and 4).

Sec. 3 is dedicated to the general consideration of the information and set hierarchy on the grounds of the information value (Eigen 1971, Volkenstein 1977), as well as on the grounds of the Russell's theory of types (Russell 1908). The concept of the information value is considered and developed.

In Sec. 4 the hierarchy among sets **S**$^{(l)}$ forming the well-ordered set **S** is considered.

In Sec. 5 the set theoretical approach to physical events and their addresses (defined as elements of two corresponding sets) representation is formulated and developed.

In Sec. 6 one defines and considers mapping with the feedback.

In Sec.7 one continues the consideration of the Sec. 4 on the setting up hierarchy inside a well-ordered set of well-ordered sets and defines the concept of microworld scales on these grounds.

In Sec. 8 indirect measurements within different scales are considered. It is indicated that possibly quarks exist not within the same scales as hadrons and leptons, but within a "smaller" scale, or exactly, within a scale of lower hierarchy.

In Sec. 9 the set theoretical concept of physical laws within a certain microworld scale is introduced and considered.

In Sec. 10 one considers consequences of the fact that a measurement within a microworld scale made by a macroscopic observer is a ***sequence*** of indirect measurements within previous larger scales. It is indicated that it leads to a fundamental limitation of our knowledge on the microworld.

In Sec. 11 the transition from quantum to the first sub-quantum scale is considered as an example of the proposed theory application.

## 2. INFORMATION TYPES AND SET TYPES (PRELIMINARY)

The purpose of the Sections 2 - 4 is to consider different ways to set up the hierarchy of sets able to treat the information: **A**) on the grounds of the theory of information by definition of the concepts of information types and the information hierarchy, and **B**) on the grounds of the Russell's theory of types (Russell 1908). It will be shown that in the case when the information and its value (Eigen 1971, Volkenstein 1977) are expressed in a language and built according logical rules the both approaches lead to the same result and that Russell's types can be expressed in terms of the value of the information obtained as result of the information treatment.

As the starting point we consider here a well-ordered finite or countable set **S** of well-ordered sets supposing that some of these sets are able to treat the information. Now introduce and set up the hierarchy among these sets *based on their properties with respect to the information treatment* and thereupon reorder set **S** with respect to hierarchy of sets forming it. The information treatment includes the following functions: 1) the receipt of information (from), 2) the sending of information (to), 3) the information processing, 4) the information interpretation, and 5) the information storage in memories. Define that the hierarchy of such a set is determined by which of these functions it executes. However, it is to take into account that this criterion could be not sufficient one to determine the hierarchy because there is the possibility that more than one set have the same type of the information treatment and, therefore, the same hierarchy according to this criterion that prevents to set up the order among them, when one reorders the set **S** with respect to hierarchies of sets forming it. It seems to be like the quantum state degeneration. This analogy suggests an idea to search for some supplementary criterions that may allow one to attribute to these sets different hierarchies (to break this "degeneration"). Then they can be ordered among themselves also with respect to their hierarchies. In the case when no such supplementary criterion exists, they can be ordered among themselves on the grounds of reasons other than hierarchy or arbitrarily.

We shall accept that the lowest hierarchy is attributed to sets that, in general, do not treat the information while the highest one is attributed to each set executed all five functions. If there is only one set of the highest hierarchy, then order (or reorder) set **S** with respect hierarchies of sets forming it so that the set possessing the highest hierarchy will be the first (we shall attribute to it № 0), while the hierarchy of other sets decreases when the number augments.

Let us reinterpret the described approach in terms of the ***information value*** (Volkenstein 1977; Eigen 1971). The information obtained by the interpretation of the processed information has the largest value because it is able to induce the most serious changes to the understanding of the obtained information meaning and, on these grounds, to invent its new applications creating material changes. For example, if to speak on physics, such an interpretation may mean the replace of existing physical laws to the new ones, which leads to a serious, maybe drastic change of our understanding of the going on in the World (*cf.* the replace of the classical mechanics to the relativity and quantum mechanics) and creation new applications, *e. g.*, nuclear energy, quantum computing *etc...* Following this way we define the hierarchy in accordance with the order of values of the information. In Sec. 3 the other approach based on Russell's theory of types (Russell 1908) will be represented.

The information on going on within a certain set of those forming the set **S** should be transmitted step-by-step to the set of the highest hierarchy to be processed and interpreted. The process of the information extraction (on occurring within a set) we shall call **measurement** considering it as a general mathematical notion. In its physical applications we, for brevity, shall use this term also for observation. For example, in the microworld one uses measurements while in the macroworld (*e. g.*, in the astronomy and astrophysics) mainly observations are used.

## 3. TYPES, INFORMATION TREATMENT AND SETS HIERARCHY (general approach)

Consider now the setting up hierarchy within the information with more details. The information is characterized by its amount (see, for example, Shannon 1948, Brillouin 1956) and value (Eigen 1971, Volkenstein 1977). Let us consider the following multi-step process:

*primary information* $\xrightarrow{creation}$ *information* $\xrightarrow{creation}$ *information* $\xrightarrow{creation}$ *information* $\xrightarrow{creation}$ *information* $\xrightarrow{creation}$ *information* $\xrightarrow{creation}$ ...

The information value can be determined as the amount of the information $I_{n+1}$ created in all $n$ these steps divided into amount of the primary information $I_1$. However, it cannot be the only characteristic of the information value because it does not take into account properties of the information content. If properties of the created information content are taken into account, then the value of the primary information should be represented by a set of such characteristics + the number defined above. Let us try to represent these characteristics in general form. Denote each of them $\xi_\nu$, where $\nu$ labels a certain characteristic. Then one has $\Xi = (\forall \nu)\{\xi_\nu\}$. Denote those of the primary information $\Xi_1 = (\forall \nu_1)\{\xi_{1,\nu_1}\}$, of the secondary information $\Xi_1 = (\forall \nu_2)\{\xi_{2,\nu_2}\}$,…, of the $n^{ts}$ step $\Xi_{n+1} = (\forall \nu_{n+1})\{\xi_{n+1,\nu_{n+1}}\}$, … *etc.*. The index at $\nu$ is necessary because for information obtained at each step the set of properties could be different from that for obtained at other steps.

Let us introduce the norm of a property $\|\xi_{l,\nu_l}\|$, which is a number. How the norm is defined depends on each concrete case, so we do not consider this problem for the general case. The complete representation of the primary information value is $\prod_{\forall l}^{\otimes} \Xi_l$, where $\Pi^\otimes$ denotes the Cartesian product of sets. Using the norm of the property one can introduce a quantity characterizing the primary information value. It is $J_V = \left(\left\|\prod_{\forall l}^{\otimes} \Xi_l\right\| + I_{n+1}\right)I_1^{-1}$ that will be called ***information value***. However, it must be kept in mind that, really, it is only a partial characteristic of the information value.

If only $n' < n$ steps are realized, while principally $n$ steps are possible, one can define the concept of the **potential information value** that is determined for all $n$ steps, no matter how many of them are realized. One can define also the concept of the **constrained information value** when constrains prohibit the realization of a part of steps.

Note that the use of the norm $\|\xi_{l,v_l}\|$ is not the only way to compare information values of primary information in different cases. It is possible to refuse from the use numbers for this purpose and instead of it to assign to each $\xi_v$ quality $\mathbf{Q}$ (which is not obligatory a number) such that between any two $\mathbf{Q}_v$ and $\mathbf{Q}_{v'}$ the relation of order, e. g., $\mathbf{Q}_{v'} \prec \mathbf{Q}_v$, exists. One can interpret this relation so that the quality $\mathbf{Q}_v$ is higher than quality $\mathbf{Q}_{v'}$. Whether this approach can be used instead of the use of $J_V = \left(\left\|\prod_{\forall l}^{\otimes} \Xi_l\right\| + I_{n+1}\right) I_1^{-1}$ to express the information value? It is possible, if relation of order like $\mathbf{Q}_{v'} \prec \mathbf{Q}_v$ can be set up between any two $\prod_{\forall l}^{\otimes} \Xi_l$ (for both cases of the initial information). But it seems questionable because different $\xi_v$ with different $\mathbf{Q}_v$ enter to this Cartesian product in complicated combinations.

***Define*** *that the type of the information is determined by its value, potential value or constrained value, accordingly to the considered problem.* ***Define*** *that the information hierarchy is set up according types of the information.*

Consider the case *when the information and its value are expressed in a language and built according the rules of logic.* Then one can define the types of information using the Bertran Russell's theory of types (Russell 1908) as the starting point. We read in the abovementioned article of Bertran Russell: "A *type* is defined as the range of significance of a propositional function, *i. e.*, as the collection of arguments for which the said function has value." "Thus whatever contains an apparent variable must be of different type from the possible values of that variable; we will say that it is of a *higher* type." In our case, for example, the processed information can be considered as the set of values of apparent variables that are, in their turn, the result of the information interpretation.

Consider it in more detail. Let $V = (\forall r[r \in \mathbf{N}; r \in [r_0, r_{max} > r_0]])\{V_r\}$ are apparent variables (Russell 1908, Whitehead & Russell 1963) of the processed, but not yet interpreted information. The interpretation consists in 1) the setting up connections between $V_r$ with different values of $r$, 2) the setting up rules how values of $V_r$ can be calculated and 3) the setting up connections with variables characterizing external factors influencing the considered system. We shall call a *theory* the result of the interpretation. $V = (\forall r[r \in \mathbf{N}; r \in [r_0, r_{max} > r_0]])\{V_r\}$ can be obtained now as values of apparent variables $U = (\forall s[s \in \mathbf{N}; s \in [s_0, s_{max} > s_0]])\{U_s\}$ of interpreted information, *i. e.*, from the theory. Therefore, according Russell the interpreted information (expressed in terms of these apparent variables) is of higher type in comparison with the processed, but not yet interpreted information. The same can be said on the not yet processed and processed information: the second is of higher type than the first one. In general, elements of a set or their configurations can be considered as values of apparent variables of the information about this set. The information has, therefore, higher type than the set itself.

Let us return to the consideration of the approach when the type of the information is defined also on the grounds of its value (Volkenstein 1977, Eigen 1971), which differs from the written above Russel's approach and is *more general because is **not** limited with the condition that the information and its value must be expressed in a language.* However, in the case when the information and it value are expressed in a language, this definition seems to be equivalent to the one based on Russell's theory of types and leads, in particular, to the same result that the interpreted information is of the highest type. Note that different levels of this interpretation may exist so that the information obtained by these kinds of interpretation could have different types. With the purpose to avoid such an uncertainty at the consideration of sets treating the information one defines the type of a set treating the information as the highest of the types of the information obtained by this treatment.

Let us now consider a well-ordered set containing sub-sets able to receive (also by performing measurements), to send, to process, to interpret and to store (in memories) the information (*cf.* Sec. 2.). We shall call such a subset **observer**, iff it is able to execute <u>all</u> these functions including measurements. We do not suppose that all considered subsets are observers, in other words, not each of them executes all the abovementioned functions.

Our purpose is to set up the hierarchy among such sets based on the information hierarchy defined above.

We define that the hierarchy order of two sets treating the information corresponds to their types order. The generalization to any finite or infinite countable well-ordered set of sets is evident. We define that *the hierarchy of a set containing a subset able to treat the information would be equal to the hierarchy of this subset*. If this set contains a set of such subsets treating the information having different types, we shall define that *the hierarchy of the considered set is determined by the highest of these types*. Thus, the type attributed by definition to subsets able to receive, process, send and store the information (they are not *observers*) would be lower than that attributed to the observer which is able also to interpret the information.

The information value probably does not affect the *original* information entropy, but **it may create the negative or positive entropy production**. For example, at the explosive crystallization of an amorphous body by laser light the information carried by this light initiates the transformation of a disordered amorphous body to the ordered crystal. The value of the original information corresponds to the big (by the absolute value) ***negative*** information entropy production. At the initiation of the explosion of an explosive by electric signal the value of the original information corresponds to the big ***positive*** entropy production because an ordered structure is turned into a disordered one. This means, the value of the original information corresponds to the absolute value of the entropy production. Define the ***specific absolute value of the entropy production*** as its absolute value divided to the original information amount. It can be an important characteristic of the information action. This connection between the information value and such a thermodynamic quantity as the entropy production suggests the idea that one can formulate the problem of the hierarchy also in terms of the thermodynamics.

Not all processes initiated by the original information are obligatory occurred at one step (*cf.* written in the beginning of this section). The "first creature" may initiate new processes creating "the second creature" *etc*. If only the "first creature" is taken into account or the following steps be prohibited by any conditions, the value of the original information would be less than in the case when the "second and following creatures" will be realized and taken into account. Therefore, the value of the information only is not enough to characterize the ability of the original information, and it is to introduce the concept of ***potential value of information*** based on taking in the account those effects that the considered information *potentially* is able to produce (maybe in some steps), but not yet produced. For example the information obtained by the observer can possess big value and potential value because its interpretation (possibly, even the creation of new physical laws) may produce remarkable effects.

Now we can define the concepts "***Value of Information***" and ***"Potential Value of Information"*** *in terms of Russell's theory of types* (Russell 1908, Whitehead & Russell 1963). We shall accept that the type of the information is determined as the type of its expression in terms of mathematical logic notions (Russell 1908).

**DEFINITION 1.** *The* **value** *of the considered information (primary information) is the highest Russell's type of the information created by the activated primary information in maximum number of executed steps.*

**DEFINITION 2.** *The* **potential value** *of the considered information (primary information) is the highest Russell's type of the information that could be created by the activated primary information in maximum number of <u>principally existing</u> steps.*

The activated information is the information that produces new information, physical, chemical, biological, industrial, social and other effects. Example: the prominent letter of Albert Einstein to Franklin D. Roosevelt, President of the USA, where Einstein proposed to begin researches aimed to create the nuclear weapon. It contained information that could be called frozen or potential one up to the moment when the President read it and decided to begin these researches. Then it became to be the active information. If the President did not read this letter or rejected the Einstein's proposal, the value of the information contained in this letter would be equal to the zero and only its potential value should be enormous.

A subset able to execute the information treatment must contain a subset formed of elements that, in their turn, are sets containing more than one element. Then in the considered subset different distributions of elements (for example, with respect to numbers of elements including to each element of this subset) can exist and, therefore, the probability and information can be defined.

The ability of such a subset to receive, send, process, store and interpret the information depends on the set structure. If there are a number of such subsets, their relative hierarchy is defined as their relative ability of the information treatment. The *rough* classification can be as follows: the lowest hierarchy (=0) have those which cannot receive, cannot send and cannot process the information; the hierarchy =1 is attributed to subsets which are able to receive, to send, to store, but cannot process the information; the hierarchy =2 is attributed to subsets which are able to receive, to send, to store and to process the information; the hierarchy =3 is attributed to subsets which are able to receive, to send, to process, to store and to interpret the information.

Inside each type could be different sub-types with different hierarchy among them. For example, inside type (3) could be different levels of the information interpretation. The highest hierarchy among them is attributed to the subset which extracts from the received and processed information general, in particular, physical laws.

Note that the active information can create new information, but it can create phenomena of different nature, *e. g.*, physical, chemical, biological, geophysical, emotions of human beings and animals, thoughts of human being expressed or not in a language, logical or not *etc.*. It must be taken into account at the consideration of the information value. The mathematical logic, in general, and Russell's theory of types, in particular, can be applied to the information value consideration only when the processes can be expressed in a language (or languages) according logical rules at all stages. Note that it must not negate without a serious consideration the possibility of existing of the information which is not expressed in a language, but despite it is built according logical rules. Of course, if it exists, these logical rules must be a *generalization* of those of the existing logic, for example, those connecting certain sets, but not propositions *etc.*.

## 4. HIERARCHY AMONG SETS $S^{(l)}$

Let there is a well-ordered not empty final or countable set **S** of not intersected not empty well-ordered sets

$\mathbf{S}^{(l)}\left[l \in \mathbf{N}, l \in [0, l_{max} \vee \infty], (l' \neq l) \Rightarrow \mathbf{S}^{(l)} \cap \mathbf{S}^{(l')} = \emptyset \vee (l' = l) \Rightarrow \mathbf{S}^{(l)}\right]$, where **N** denotes the set of all natural numbers. The written above allows one to set up the hierarchy between all well-ordered sets forming the well-ordered set **S**.

One can order the set $\mathbf{S} = \{\mathbf{S}^{(l)}\}$ with respect hierarchies of sets $\mathbf{S}^{(l)}$. Let us set up the hierarchy within the set $\mathbf{S}$ so that the set $\mathbf{S}^{(0)}$ possesses the highest hierarchy and the hierarchy of sets $\mathbf{S}^{(l)}$ decreases with the increase of $l$: $l' > l \Rightarrow \tilde{h}\mathbf{S}^{(l')} < \tilde{h}\mathbf{S}^{(l)}$, where $\tilde{h}$ denotes *hierarchy*. Let us postulate that only the set $\mathbf{S}^{(0)}$ is allowed to receive, to send, to process, to interpret and to store the information on all other sets. This means, we consider here the case when the observer(s) exists at only one scale, namely that having the highest hierarchy. At applications to microworld physics (see below) this means that there is only macroscopic observer(s). The corresponding situation arising at the macroworld study merits a special consideration; the consideration of the microworld (see below) cannot be automatically transferred to the macroworld. The information on events is provided by a certain mathematical procedure that we shall call *measurement* or *observation*. The use of these two terms is dictated by some applications of this mathematical theory: for example, in the microworld usually one uses measurements while in the astronomy and astrophysics observations are usually used to provide and obtain information. However, for brevity, we shall use the term measurement in both cases, keeping in mind that it includes also observation. The set $\mathbf{S}^{(0)}$ contains a subset $U_{\mathbf{S}^{(0)}} \subset \mathbf{S}^{(0)}$ possessing the following properties: it is able to receive (also *by measurements*), to process, to interpret, and to store the information. We shall call this subset $U_{\mathbf{S}^{(0)}}$ **observer**. The abovementioned interpretation is done on the grounds of certain laws (mathematical, physical *etc.*) that should be expressed in a convenient mathematical form and included to the subset $U_{\mathbf{S}^{(0)}}$. If a certain deviation $\delta\mathbf{S}^{(l')}\left(l'\in\{l\}, l'>0\right)$ has occurred with the set $\mathbf{S}^{(l')}$ itself, the information on it must be forwarded step-by-step to the subset $U_{\mathbf{S}^{(0)}}$ to be processed, interpreted and stored.

# 5. SETS OF EVENTS AND THEIR ADDRESSES

Let there are two not empty well-ordered sets (see, for example, Jech 2003) $A=\{a\}$ of elements $a$ that we shall call **events** and $\overset{(*)}{H}=\{h\}$ of elements $h$ that we shall call the **set of addresses** of elements $a$.

Consider a not empty subset $Y\subseteq A$. Set up the homomorphism keeping the order between $Y$ and $\overset{(*)}{H}_Y\subseteq\overset{(*)}{H}$, where $\overset{(*)}{H}_Y$ be homomorphic map of $Y\subseteq A$. We shall call $\overset{(*)}{H}_Y$ the address of the subset $Y$ of events.

In particular, if $Y=a$, then $\overset{(*)}{H}_Y=\overset{(*)}{H}_a$ will be the address of the single event $a$. We use the homomorphic, but not the isomorphic mapping, taking into account that more than one event may have the same address.

Consider the following case

$$\left(\exists \overset{(*)}{H}, \exists H, (\forall i)\left[\exists \overset{(i)}{H}\right] (\forall i, \forall n)[n \wedge i \in \{\mathbf{N}\}]; \forall i \in [1, n]\right)\left[\overset{(*)}{H} \supseteq H = \prod_{i=1}^{n} {}^{\otimes} \overset{(i)}{H}\right], \quad (1)$$

where $\{\mathbf{N}\}$ is the set of all natural numbers, $\overset{(*)}{H}, H$ and all $\overset{(i)}{H}$ are well-ordered sets.

Let us consider the following particular cases:

1. If $H = \overset{(*)}{H}$, the address of a subset $Y$ of events can be represented as

$$\overset{(*)}{H}_Y = H_Y = \prod_{i=1}^{n} {}^{\otimes} \overset{(i)}{H}_Y, \quad (2)$$

2. while if $H \subset \overset{(*)}{H}$, the following three cases are possible:

$$\mathbf{a}) \overset{(*)}{H}_Y \subset H, \mathbf{b}) \overset{(*)}{H}_Y \subset M = \overset{(*)}{H} \setminus H, \mathbf{c}) \overset{(*)}{H}_Y \cap H \neq \emptyset \wedge \overset{(*)}{H}_Y \cap M \neq \emptyset \quad (3)$$

In the case (**a**) Eqn. (2) is valid. However, the cases (**b**) and (**c**) demand special considerations.

We see two options. **Option I**: An expansion of the set $\overset{(*)}{H}$ so that the obtained new set $\overset{(**)}{H} \supset \overset{(*)}{H}$ would be represented in the form

$$((\forall i, \forall m)[m \wedge n \wedge i \in \{\mathbf{N}\}])[i \in [1, m > n]]\left[\exists \overset{(i)}{H}, \overset{(**)}{H} = \prod_{i=1}^{m} {}^{\otimes} \overset{(i)}{H}\right], \quad (4)$$

therefore, addresses will be considered as subsets of $\overset{(**)}{H}$, i. e., as $\overset{(**)}{H}_Y \subseteq \overset{(**)}{H}$.

**Option II**: That in two of abovementioned cases the address of $Y \subseteq A$

cannot be represented by Eqn. (2) and should be remained in the form $\overset{(*)}{H}_Y$. The property of the ordering of the set $\overset{(*)}{H}$ allows one to write:

$$\left(\exists \overset{(*)}{H}_1 \subset \overset{(*)}{H}, \not\exists \overset{(*)}{H}_2 \subset \overset{(*)}{H} \vee \left(\left(\exists \overset{(*)}{H}_2 \subset \overset{(*)}{H}\right)\left[\mu\left(\overset{(*)}{H}_2\right)=\emptyset\right]\right)\right)$$
$$\left[\overset{(*)}{H}_1 \subset \overset{(*)}{H}, \overset{(*)}{H}_1 \prec \overset{(*)}{H}_Y, \overset{(*)}{H}_1 \prec \overset{(*)}{H}_2 \prec \overset{(*)}{H}_Y\right] \qquad (5)$$

and

$$\left(\exists \overset{(*)}{H}_3 \subset \overset{(*)}{H}, \not\exists \overset{(*)}{H}_4 \subset \overset{(*)}{H} \vee \left(\left(\exists \overset{(*)}{H}_4 \subset \overset{(*)}{H}\right)\left[\mu\left(\overset{(*)}{H}_4\right)=\emptyset\right]\right)\right)$$
$$\left[\overset{(*)}{H}_3 \subset \overset{(*)}{H}, \overset{(*)}{H}_3 \succ \overset{(*)}{H}_Y, \overset{(*)}{H}_3 \succ \overset{(*)}{H}_4 \succ \overset{(*)}{H}_Y\right], \qquad (6)$$

where $\mu(H)$ denotes the measure of the set H.

## 6. MAPPING WITH FEEDBACK

We wrote above on a mapping (homomorphism) of set $A$ subsets to set $\overset{(*)}{H}$ subsets. For the application of this formalism to a real system, e. g., physical system, a certain real procedure is necessary 1) to establish the demanded correspondences and 2) to make the result, *i. e.*, the address, known. The latter problem will be considered in this Section.

Let us consider a simple example. The set of all apartments in a building is mapped to a subset of the set of all natural numbers so that each apartment has its number. But usually this number is put on the apartment door, in other words, the apartment is labeled by its number.

Then the number of the apartment, *i. e.*, the result of the abovementioned mapping, becomes known.

On the analogy of this example let us consider now how the address $H_Y^{(*)}$ can be found out, in other words, by means of what the address of the subset of events $Y \subseteq A$ will become known. The ordinary procedure of the direct mapping of $Y \subseteq A$ to $H_Y^{(*)}$ does not undertakes this task. At the same time several applications of the mapping, for example, measuring in physics, include this task and so a corresponding mathematical procedure is needed to accomplish it. Indeed, a measurement of space - time co-ordinates of an event would be useless, if the observer cannot obtain its result, in other words, if there is not feedback between the sending a (light) signal for a measurement and obtained results of it. That is why we want to label event by its address. It allows the observer at one go to get to know the event and its address. Of course, label may be changed from measurement to measurement. We shall call such an event (an element of the set *A*) with the label a ***labeled event (a labeled element of the set A)***. Probably, this labeling would be important also at the study of the macro-World (the Universe and its regions), but this is not a subject of the present paper.

Now introduce the necessary feedback procedure. Let each element $a \in A$ and each element $h \in H^{(*)}$ are themselves sets of two or more elements (this is the necessary condition that the feedback is possible):

$$a = (\forall n \in \mathbf{N}, \forall j \in [1, n]) \begin{Bmatrix} \tilde{a} \\ b_{a,j} \end{Bmatrix}, \tag{7}$$

$$h = (\forall m \in \mathbf{N}, \forall l \in [1, m]) \begin{Bmatrix} \tilde{h} \\ q_{h,l} \end{Bmatrix}, \tag{8}$$

where $\tilde{a}$ and $\tilde{h}$ denote the event and the address themselves, while $b_{a,j}$ and $q_{h,l}$ are intermediate objects using for the following mapping procedure. In the beginning for the sake of simplicity we shall consider the case when $n=1$ and $m=1$, i. e., when there is only one $b_{a,1} = b_a$ and only one $q_{h,1} = q_h$. Map now $a \in A$ to an element $h \in \overset{(*)}{H}$. Then we shall obtain the pair $b_a q_h$. Now one can map this pair to the corresponding element $a$ of the set $A$ (according the index $a$ of $b_a$). This means, we shall return element $b_a$ to their place, but together with corresponding label $q_h$. Thus, now an element $a$ of the set $A$ is labeled by its address $q_h$. The presence of $b_a$ in the pair $b_a q_h$ establishes the isomorphism between $\{b_a q_h\}$ and $\{a\}$ in the situation when the mapping $\{a\}$ to $\{h\}$ is a homomorphism, and, therefore, the inverse mapping made directly as $\{h\} \to \{a\}$ would be not single-valued.

The considered situation is like the one arising in the quantum mechanics in the case of quantum state degeneration. Then a new factor breaking the symmetry, for example, magnetic field, can remove the degeneration splitting the degenerated energy level into a number of closed different levels. In our case this role plays $b_{a,j}$ and $q_{h,l}$ that turn the homomorphism into isomorphism (of course, not for $\tilde{a}$ and $\tilde{h}$, but for $a$ and $h$ defined by Eqns. (7) and (8), correspondingly).

The procedure described above can be done with each element $a \in Y \subseteq A$, and we shall obtain

$$Y = \left\{ a = (\forall j \in [1,n]) \begin{Bmatrix} \tilde{a} \\ b_a q_h \end{Bmatrix} \right\} \tag{9}$$

Thus, the described procedure establishes the necessary feedback labeling the subset $Y$ by the addresses of its elements. We shall call this mathematical procedure **measurement**, though it could not be obligatory the measurement in the physical meaning.

## 7. SETS' HIERARCHY

Let there is a countable (finite or infinite) well ordered set $\mathbf{S} = \{\mathbf{S}^{(q)}\}$ of finite or infinite (countable or continuum) well ordered in pairs non-intersected sets $\mathbf{S}^{(0)}, \mathbf{S}^{(1)}, \mathbf{S}^{(2)}, \ldots, \mathbf{S}^{(q)}, \ldots$,

$(\forall (q,q'))[\mathbf{S}^{(q)} \cap \mathbf{S}^{(q')}] = \emptyset$, where $q \in [0, q_{max} \vee \infty]$, $q_{max} \in \mathbf{N}, q_{max} \geq 1$.

In the pure mathematical framework the order within the set $\mathbf{S} = \{\mathbf{S}^{(q)}\}$ can be set up, for example, according the *order of types* of $\forall \mathbf{S}^{(q)}$, or by another way. However, for the applications to physical problems the order in the set $\mathbf{S} = \{\mathbf{S}^{(q)}\}$ is to be set up *on the grounds of physical* *reasons*, including inferences arising from experimental results. If this order be set up on the grounds of mathematical reasons only, possibly the obtained mathematical theory would not be fit for the considered physical problems treatment.

Let the set **S** is reordered so that the hierarchy among the sets $\mathbf{S}^{(q)}$ corresponds to their order such that the highest hierarchy is attributed to the set $\mathbf{S}^{(0)}$. Define 1) that the ability to initiate indirect measurements in all $\mathbf{S}^{(q)}$ is attributed only to the set $\mathbf{S}^{(0)}$, and 2) that the information obtained from all such measurements **can be extracted only from** $\overset{(*)}{H}{}^{(0)}$**, but not from any** $\overset{(*)}{H}{}^{(q>0)}$**.**

This consideration allows one to define the notion **SCALE OF THE MICROWORLD** as follows: *we shall call the set* $\mathbf{S}^{(q)}$ ***scale*** *number q (q=0,1,2,3,...) of the microworld when namely the microworld is studied and number q=0 corresponds to the macroscopic scale. The observer is always macroscopic and makes measurements within the scale* $\mathbf{S}^{(0)}$.

## 8. MEASUREMENTS

In physics the information on a physical object is obtained by measurements. We shall keep this term also for the case of mathematical objects. Consider this problem in detail.

Measurement in physics can be performed ***directly*** by a <u>macroscopic</u> observer (human being or automaton) using measuring instruments. The task of the observer includes the interpretation of measurement results. Because the observer is *always* macroscopic, his ability to make **direct measurements** is limited with the atomic and nuclear scale. It is questionable whether they can be used within smaller scales, if they exist. Maybe it is possible because the development of particle accelerators to the direction of higher-and-higher energies, but, as it can be concluded from our consideration, the effective use of such measurement equipments is possible only in combination of direct and indirect measurements.

Mandelstam (1972) introduced the concept of **indirect measurement** (see also Braginsky & Khalili (1992), Auletta (2000) ) to quantum mechanics. We shall try to use indirect measurements to penetrate step-by-step into smaller and smaller scales of the microworld, precisely speaking, into scales of lower and lower hierarchy. The general theory of indirect measurements is developed in Sec. 3.4* of the book (Braginsky & Khalili (1992)). However, this theory supposes that all systems participating in the indirect measurement process, in exception of the macroscopic observer, are quantum ones. In the present work we shall consider the penetration to the sub-quantum scale and beyond. There is no reason *à priori* to suppose systems within such scales to be also quantum. Therefore, the indirect measurement theory of Braginsky & Khalili (1992)) cannot be applied to the cases that are the subject of the present work, and so we develop here another theory of indirect measurements.

The application of Mandelstam's idea for this purpose can be represented by the following example. The macroscopic observer measures co-ordinate of a particle $Q$ within the atomic scale. Thereupon this particle collides with an object belonging to the nuclear scale, and the macroscopic observer measures the change of $Q$'s co-ordinate. The comparison of these two measurements provides information on the nuclear scale object.

Consider this problem firstly as the pure mathematical one. Let results of two subsequent direct measurements are different so that their difference exceeds the statistical error of measurement. This means that a subset $Y \subseteq A$ of events has two "addresses" $\overset{(*)}{H}_{Y,1}$ and $\overset{(*)}{H}_{Y,2}$ such that

$\overset{(*)}{H}_{Y,1} \prec \overset{(*)}{H}_{Y,2}$ , $\overset{(*)}{H}_{Y,1} \cap \overset{(*)}{H}_{Y,2} = \emptyset$. This fact may have different interpretations. Possibly, something was changed by itself in the system $Y \subseteq A$ between these two measurements, if they were not made simultaneously. If it is proved that such a possibility does not exist in this case, then one of the remained possibilities is that there is one more subset $Y' \subseteq A$ of events, the "address" of which we shall denote $\overset{(*)}{H}_{Y'}$, that influences the address of the subset $Y$ so that $\exists \mathbf{F}\left(\overset{*}{H}_{Y,1}, \overset{*}{H}_{Y,2}\right) = \overset{*}{H}_{Y'}$. This equality means that this pair of addresses $\overset{(*)}{H}_{Y,1}$ and $\overset{(*)}{H}_{Y,2}$ contains the information sufficient to define the address $\overset{(*)}{H}_{Y'}$ of a subset $Y'$ of events. The function $\mathbf{F}\left( \overset{(*)}{H}_{Y,1}, \overset{(*)}{H}_{Y,2} \right)$ means a homomorphism of the pair $\overset{(*)}{H}_{Y,1} \subset \overset{(*)}{H}$ and $\overset{(*)}{H}_{Y,2} \subset \overset{(*)}{H}$ to a subset $\overset{(*)}{H}_{Y'} \subset \overset{(*)}{H}$. However, it is possible that the abovementioned pair of addresses does not contain the sufficient information on the address $\overset{(*)}{H}_{Y'}$, which means that $\nexists \mathbf{F}\left( \overset{*}{H}_{Y,1}, \overset{*}{H}_{Y,2} \right) = \overset{*}{H}_{Y'}$. Consider now the general case when the number of measurements is not limited with two. Define the function

$$\mathbf{F}_m\left( \overset{(*)}{H}_{Y,1} \prec \overset{(*)}{H}_{Y,2} \prec \cdots \prec \overset{(*)}{H}_{Y,m}; (\forall i)\left[ \overset{(*)}{H}_{Y,i+1} \bigcap_{i=1}^{m} \overset{(*)}{H}_{Y,i} = \emptyset \right] \right), \tag{10}$$

where $m \geq 1$. Denote by $\alpha$ type of subset of events. Then $\overset{(*)}{H}_{Y'_\alpha}$ is defined as the address of a subset $Y'_\alpha$ of events

$$\Leftrightarrow \left\{ \left( \exists \left\{ \overset{(*)}{H}_{Y,i} \right\}^{(\alpha)} \right) \left[ \exists \left( \lim_{m_\alpha \to \infty} \mathbf{F}_{m_\alpha}^{(\alpha)} = \overset{(*)}{H}_{Y'_\alpha} \neq \emptyset \right) \wedge \nexists \left( \lim_{m_\alpha \to \infty} \mathbf{F}_{m_\alpha}^{(\alpha)} = \overset{(*)}{H}_c \neq \overset{(*)}{H}_{Y'_\alpha} \right) \right] \right\}, \tag{11}$$

where $\exists \left\{ \overset{(*)}{H}_{Y,i} \right\}^{(\alpha)}$ means, "Exists set (sequence) $\{ \}^{(\alpha)}$".

It is evident from Eqn. (11) and $Y'_\alpha$ definition that 1) $\overset{(*)}{H}_{Y'_\alpha}$ cannot be detected directly, but only indirectly by its influence upon the subset $Y$ of events, and 2) $Y'_\alpha$ itself cannot be detected by this way, but only its address.

Now let $Y \overset{def}{=} Y^{(q)} \subseteq \mathbf{S}^{(q)}$. Then $Y' \overset{def}{=} Y^{(q+1)} \subseteq \mathbf{S}^{(q+1)}$. If in all considerations and formulas above one replaces $Y^{(q)}$ to $Y^{(q+1)}$, and $Y^{(q+2)}$ to $Y^{(q+3)}$, one obtains the address of a subset $Y^{(q+2)}$ of events. The following step will be evidently the replace $Y^{(q+1)}$ to $Y^{(q+2)}$ and $Y^{(q+2)}$ to $Y^{(q+3)}$. Thus, the address of a new subset of events $Y^{(q+3)}$ will be obtained. This procedure can be continued.

The purpose of this Section is to find a way to detect events within a scale $(q+1)$ by use of indirect measurements made within the scale $q$. It will be one step of the multi-step indirect measurement made by an observer who always is macroscopic, *i. e.*, being and making measurements within the scale $q = 0$. Such a measurement consists of subsequent applications of described one step indirect measurements (that we denote $q' \to q'+1$) beginning from that $0 \to 1$ up to the desired measurement $q \to q+1$. Now rewrite Eqns. (10) and (11) for a measurement $q \to q+1$:

$$\mathsf{F}_m^{(q)}\left( \overset{(*)}{H}_{Y^{(q)},1} \prec \overset{(*)}{H}_{Y^{(q)},2} \prec \cdots \prec \overset{(*)}{H}_{Y^{(q)},m} ; (\forall i)\left[ \overset{(*)}{H}_{Y^{(q)},i+1} \bigcap_{i=1}^m \overset{(*)}{H}_{Y^{(q)},i} = \emptyset \right] \right) \tag{12}$$

Then $\overset{(*)}{H}{}^{(q+1)}_{Y^{(q+1)},\alpha}$ is defined as the address of a subset $Y^{(q+1)}_{Y^{(q+1)},\alpha}$ of events within the scale $\mathsf{S}^{(q+1)}$

$$\Leftrightarrow \left\{ \begin{matrix} \left( \exists \left\{ \overset{(*)}{H}{}^{(q|\alpha)}_{Y^{(q)},i} \right\}^{(q|\alpha)} \right) \\ \left[ \left( \exists \lim_{m_{q|\alpha} \to \infty} \mathsf{F}^{(q|\alpha)}_{m_\alpha} = \overset{(*)}{H}{}^{(q+1)}_{Y^{(q+1)},\alpha} \neq \emptyset \right) \wedge \not\exists \left( \lim_{m_{q|\alpha} \to \infty} \mathsf{F}^{(q|\alpha)}_{m_{q|\alpha}} = \overset{(*)}{H}_c \neq \overset{(*)}{H}{}^{(q+1)}_{Y^{(q+1)},\alpha} \right) \right] \end{matrix} \right\} \tag{13}$$

It is important to find how many and what namely independent data are included to each of these addresses. The answer to these questions establishes the physical grounds within each scale and by this indicates what is to be determined by experiments to study physical events within a certain scale.

Let the Eqn. (13) is not satisfied. It is possible that it is induced by events taking place within a set $\mathbf{S}^{(q+2)}$, the set (scale) which's existence was not yet known. If the Eqn. (13) be satisfied by the substitution scale of number $(q+1)$ as the smallest hierarchy scale to the one of number $(q+2)$, then this hypothesis will be confirmed. In this case the measurement is to be done firstly within the scale $\mathbf{S}^{(q+1)}$ from the one $\mathbf{S}^{(q)}$, but it must measure the address of an event for which the Eqn. (13) is satisfied. Then, as the second step, it is to make measurement within the scale $\mathbf{S}^{(q+2)}$ from the one $\mathbf{S}^{(q+1)}$ of the address of the event which is responsible for the phenomenon mentioned above, *i. e.*, that the Eqn. (13) was not satisfied.

Let $q=0$, and the measurements are made by a macroscopic observer. According to the written above he is able to measure not only within the scale $\mathbf{S}^{(1)}$, but also within the scale $\mathbf{S}^{(2)}$ and, apparently, beyond because the procedure described above can be continued to $q>2$. This means, if we begin from the macroscopic observer, *i. e.*, from $\mathbf{S}^{(0)}$, this procedure opens him the way to make measures within smaller and smaller scales. According to the written above this penetration to smaller and smaller scales is realized if at each step, in exception of the last one, Eqn. (13) is not satisfied and, <u>*therefore, the result exists only at the last step of these multi-step measurement.*</u>

It could happen that results of measurements of taking place within a scale number $(q)$ can be interpreted on the grounds of the hypothesis that they are produced by the existence of some new objects that were not detected in this scale. However, it is possible that these hypothetic objects are within the scale $(q+1)$, but not $(q)$. Possibly, they can be detected there <u>by direct measurements</u> using up to date techniques, *e. g.*, accelerators producing extremely high energy particles. Then their direct detection will confirm the abovementioned hypothesis. Perhaps, this is namely the situation with quarks (see, for example, Kokkedee 1969, Nambu 1985, Gribov 2002, Hosaka&Toki 2001, Ne'eman 1961, Gell-Mann&Ne'eman 1964): <u>***quarks exist not within the same scale where other elementary particle exist, but within the following ("smaller", exactly, of lower hierarchy) scale.***</u> Among while this interpretation leads to the following question: whether the space-time continuum exists within the very "small" quark scale? It is very probably that the answer is negative. If so, notions of transformations of (non-existing) co-ordinate systems and corresponding groups are nonsense within the quark scale.

## 9. PHYSICAL LAWS WITHIN EACH SCALE

Represent in the general form physical laws within a certain scale $\mathbf{S}^{(q)}$. Let measurements have determined the address $\overset{(*)\,(q)}{H}_{Y^{(q)}}$ of the subset $Y^{(q)} \subseteq A^{(q)}$. Introduce a set $\{\mu_{Y^{(q)}}^{(q)}\} \overset{def}{=} M_{Y^{(q)}}^{(q)} \overset{def}{=} M^{(q)}(Y^{(q)} \subseteq A^{(q)})$ that attributes certain properties to the set $\overset{(*)\,(q)}{H}_{Y^{(q)}}$, which means, in particular, *the choice of the interpretation of the obtained measurements' results.*

Call $M_{Y^{(q)}}^{(q)} \overset{def}{=} M^{(q)}(Y^{(q)} \subseteq A^{(q)})$ **associated set** of the subset $Y^{(q)} \subseteq A^{(q)}$ of events. The interpretation of address measurement results for different subsets $Y^{(q)} \subseteq A^{(q)}$, i. e., different $M^{(q)}(Y^{(q)} \subseteq A^{(q)})$, may be interdependent. It suggests an idea to introduce the set associated with the set of all subsets of the set of events:

$$\{\overset{*}{\mu}{}^{(q)}\} \overset{def}{=} \overset{*}{M}{}^{(q)} = (\forall Y^{(q)} \subseteq A^{(q)})[\{M^{(q)}(Y^{(q)} \subseteq A^{(q)})\}] \tag{14}$$

The set $\overset{*}{M}{}^{(q)}$ we shall call **the complete physical theory within the scale $\mathbf{S}^{(q)}$**, while $M^{(q)}(Y^{(q)} \subseteq A^{(q)})$ we shall call **a partial physical theory within the scale $\mathbf{S}^{(q)}$**. The choice of sets $\overset{*}{M}{}^{(q)}$ and $M^{(q)}(Y^{(q)} \subseteq A^{(q)})$ really means the introduction of **models** because there is a certain freedom of their choice, but not a "*categorical imperative*" what namely is to be chosen as the theory.

One of possible options is to choose $M^{(q)}(Y^{(q)} \subseteq A^{(q)})$ and $\overset{*}{M}^{(q)}$ as sets of *operators* $\{\hat{u}_{Y^{(q)}}^{(q)}\} \overset{def}{=} \hat{M}^{(q)}(Y^{(q)} \subseteq A^{(q)})$ and

$$\left\{\overset{\hat{*}}{\mu}\right\} \overset{def}{=} \overset{\hat{*}}{M}^{(q)} = (\forall Y^{(q)} \subseteq A^{(q)})\left[\left\{M^{(q)}(Y^{(q)} \subseteq A^{(q)})\right\}\right]$$ over the set $H^{(*)(q)}$. How these operators determine the interpretation of measurements' results? If $H^{(*)(q)}$ is a space, they can map (project) the set $H^{(*)(q)}$ or its subsets to another space $R^{(q)}$, *e. g.*, a Hilbert space, and its subsets. It can be written as follows:

$$\hat{M}^{(q)}(Y^{(q)} \subseteq A^{(q)})\left(H_{Y^{(q)}}^{(*)(q)} \subseteq H^{(*)(q)}\right) = R_{Y^{(q)}}^{(q)} \subseteq R^{(q)} \qquad (15)$$

The measurement results are none other than the set of elements $\left(H_{Y^{(q)},mes}^{(*)(q)} \subseteq H^{(*)(q)}\right)$. These results can be interpreted only after the following operator $\hat{M}_{Y^{(q)}}^{(q)}$ action:

$$\hat{M}_{Y^{(q)}}^{(q)}\left(H_{Y^{(q)},mes}^{(*)(q)} \subseteq H^{(*)(q)}\right) = R_{Y^{(q)},mes}^{(q)}, \qquad (16)$$

*i. e.*, not the obtained results themselves, but their projection to the space $R^{(q)}$ should be used for the interpretation.

Thus, the transition from $q$ to $q+1$ means the search for convenient models (=physical theories) for interpretation of results of indirect measurements made on the scale $\mathbf{S}^{(q)}$ to determine addresses of events on the scale $\mathbf{S}^{(q+1)}$ under the condition that the physical laws on the scale $\mathbf{S}^{(q)}$ are already known. These laws are necessary for the theory of the abovementioned indirect measurements (*cf.* Braginsky & Khalili (1992), Auletta (2000)). Without them the interpretation of such indirect measurements would be impossible.

Let us use the analogy with the approach to classical scale – quantum scale transition. It must keep in mind that the choice of this way is of the **hypothetical** character, and that other, probably also hypothetical ways may exist to be used for the construction of physical theories within different scales. Following this way one could replace physical quantities, *i. e.*, **the addresses that principally can be measured**, defined within the scale number $q$ to operators within the scale number $(q+1)$, which reminds the transition from classical to quantum scale. The obtained operators just exactly will form the theory $M^{\hat{*}\,(q+1)}$.

The set $\{h^{(q+1)}\} = H^{(*)(q+1)}$ may, in particular, be Hilbert space or its subset, upon which act operators of the theory $M^{\hat{*}\,(q+1)}$. This reminds the approach of the quantum mechanics where quantities of the classical mechanics are replaced by operators acting upon probability amplitudes. Try to keep this way of $q=0$ to $q=1$ transition for all values of $q$. In quantum mechanics the wave function is the function of all the addresses of the classical mechanics, *e. g.*, co-ordinate, or linear momentum, and time. So we shall accept that $(\forall q)$ addresses $h^{(q+1)} \in \overset{(*)}{H}{}^{(q+1)}$ would be "wave functions" (probability amplitudes) of addresses $h^{(q+1)}$. If one accepts this way, there is no reason to keep the classical notion of the address used within the quantum and other scales. So we shall consider $(\forall q)[h^{(q)}]$ <u>as the address itself</u>. By this definition we break the <u>direct</u> connection between a measurement result and the address because, for example, in quantum mechanics the wave function is the amplitude of probability of addresses (if to stay on the probabilistic interpretation of quantum mechanics). This means, if to accept this definition, the address is not obligatory measurable, but it can serve for the interpretation of measurement results ***in the framework of a certain theory***, as it is being done in quantum mechanics.

## 10. MORE ON MACROSCOPIC OBSERVER MEASUREMENT WITHIN A SCALE $S^{(q+1)}$

It was considered above (Sec. 5) how a scale $(q+1)$ can be detected by **indirect measurements with feedback** within the scale $q$ and how it can be continued to scales $q+2$ and beyond. Now, taking into account the written in Sec. 9, we can give more concrete expression to this procedure. These measurements may detect that there is an event or a set of events violating the laws $\overset{\hat{*}}{M}{}^{(q)}$. Then, one of possible ways to interpret this fact would be the assumption that a smaller (of smaller hierarchy) scale $\mathbf{S}^{(q+1)} < \mathbf{S}^{(q)}$ exists with its own physical laws $\hat{\overset{*}{M}}{}^{(q+1)}$. The task now is to find $\hat{\overset{*}{M}}{}^{(q+1)}$. Remind that the *multi-step indirect measurement* is a **sequence** of indirect measurements (*cf.* the end of Sec. 5) realized, as it was describing above, steps from $\mathbf{S}^{(0)}$ to $\mathbf{S}^{(1)}$, from $\mathbf{S}^{(1)}$ to $\mathbf{S}^{(2)}$, from $\mathbf{S}^{(2)}$ to $\mathbf{S}^{(3)}$, ..., from $\mathbf{S}^{(q)}$ to $\mathbf{S}^{(q+1)}$. All these measurements are made in succession *by a macroscopic observer* who himself, by the definition, is within the scale $\mathbf{S}^{(0)}$. Such a multi-step measurement can be represented by the following scheme.

1. The macroscopic observer finds out that results of certain measurements cannot be interpreted on the grounds of physical laws $\hat{\overset{*}{M}}{}^{(0)}$ within the scale $\mathbf{S}^{(0)}$ because Eqn. (13) is not satisfied in this case.

2. He finds out that no change of these physical laws can change this fact. Then he supposes that the scale $\mathbf{S}^{(1)}$ of lower than $\mathbf{S}^{(0)}$ hierarchy exists with its physical laws $\hat{\overset{*}{M}}{}^{(1)}$.

3. Then the macroscopic observer finds out that results of some of his indirect measurements made in the scale $\mathbf{S}^{(0)}$ to study occurring within the scale $\mathbf{S}^{(1)}$ cannot be interpreted on the grounds of physical laws $\hat{\overset{*}{M}}{}^{(1)}$ or any other physical laws within this scale, and supposes that the scale $\mathbf{S}^{(2)}$ of lower than $\mathbf{S}^{(1)}$ hierarchy exists with its physical laws $\hat{\overset{*}{M}}{}^{(2)}$.

4. *Etc.*

**<u>Note</u>**. It is important to remind that at each step the set $\hat{\overset{*}{M}}{}^{(q)}$ is *not the only possible* physical laws. Therefore, those measurements' results that cannot be interpreted on the grounds of the physical laws $\hat{\overset{*}{M}}{}^{(q)}$ may be interpreted on the grounds of the other physical laws $\hat{\overset{*}{M}}'{}^{(q)}$ within the same scale $\mathbf{S}^{(q)}$ without hypothesis on the $(q+1)$-th scale existence. Only if it be found impossible, the existence of a new scale of lower hierarchy can be supposed and considered. Note that it demands to be very careful because results of some other kinds of measurements described well by $\hat{\overset{*}{M}}{}^{(q)}$ may be incompatible with $\hat{\overset{*}{M}}'{}^{(q)}$.

Each measurement of event addresses within the scale $(q+1)$ performed by a macroscopic observer is a multi-step sequence of measurements. This fact is a matter of principal. Indeed, this sequence of measurements with the feedback (see Sec. 3) at each step demands a certain time $\vartheta_{0,q+1}$ *determined by the observer's clock* which is apparently an increasing function of the number of steps. Note that it is necessary to use measurements with the feedback, so the time $\vartheta_{0,q+1}$ includes the time of measurement itself and the feedback time.

In view of this it could be expected that the maximum value $\max q \stackrel{def}{=} q_{\max}$ exists that limits our sequential penetration into smaller and smaller microworld scales. The reason of such a limitation existence is that there is the maximum permitted time, $\max \vartheta_{0,q+1} = \vartheta_{0,q+1,\max}$ of measurement still allowing the observer to attribute a certain time moment (with a *reasonable* error) to the information provided by the measurement within the scale $\mathbf{S}^{(q+1)}$, while for $\vartheta_{0,q+2} > \vartheta_{0,q+1,\max}$ it becomes impossible. Even if we shall refuse from a dynamic description of the event behavior, *i. e.*, from its description as function of time, and will limit ourselves with only the connection between the initial and final states (the basic idea of the S-matrix method in the quantum collision theory), *physically* the time interval between the initial state creation and the appearance of the final one really cannot be ∞, but must be finite. This demands issues from the necessity to avoid processes other than the studied one to occur during this time interval (remind: observer's clock!), simply, not to blend different processes in our consideration. One more argument exits in favor of this limitation. The information that can be provided by a multi-step indirect measurement made within the scale $\mathbf{S}^{(q+1)}$ would be expected to be of small amount and scant as to its content in comparison with that provided by measurements made within the scale $\mathbf{S}^{(q)}$. Indeed, for example, measurements within the scale $\mathbf{S}^{(1)}$ provide information really scant as to its content in comparison with those made within the scale $\mathbf{S}^{(0)}$: within the classical scale $\mathbf{S}^{(0)}$ one can measure simultaneously particle co-ordinate and corresponding linear momentum, while within the quantum scale $\mathbf{S}^{(1)}$ it is impossible. It would be natural to suppose that this effect occurs at each $q' \to (q'+1)$. Then the considered effects $\forall q' \in [0, q+1]$ contribute to the resulting one for the information on events' addresses within the scale $\mathbf{S}^{(q+1)}$ obtained by the macroscopic observer.

Note that this fundamental limitation of the possibility of knowing the microworld and its physical laws creates the following problem. Let within the scale $\mathbf{S}^{(q_{max}+1)}$ an event or a subset of events exist violating physical laws $\hat{M}^{(q_{max}+1)}$ of this scale. In this case scales $\mathbf{S}^{(q_{max}+2)}$ and smaller do not exist. Therefore, this effect cannot be created by events within a neighbor smaller scale (*cf.* the written above). Then, what is its nature and origin? One way to eliminate this problem could be a kind of renormalization, which means a relevant change of physical laws, for example, the replacement of $\hat{M}^{(q_{max}+1)}$ to another $\hat{M}'^{(q_{max}+1)}$, but it is not always possible, as it was indicated above.

## 11. QUANTUM → SUB-QUANTUM SCALE TRANSITION

We have denoted classical scale $\mathbf{S}^{(0)}$ and quantum scale $\mathbf{S}^{(1)}$. Consider the penetration to the closest sub-quantum scale $\mathbf{S}^{(2)}$ starting from $\mathbf{S}^{(1)}$. The set $H^{(*)}_{(q)}$ at $q = 1$ is the Hilbert space of states of a quantum system $Y^{(1)} \subseteq A^{(1)}$. Then its address will be $H^{(*)(1)}_{Y^{(1)}}$ that can be wave function or density matrix defined over the set $A^{(1)}$.

At the transition from classical scale $\mathbf{S}^{(0)}$ to the quantum scale $\mathbf{S}^{(1)}$ one obtains addresses in the form of wave functions defined over space-time continuum instead those in the form of space-time points. Now we transit from $\mathbf{S}^{(1)}$ to $\mathbf{S}^{(2)}$. Of course, different versions are possible in this case, but in the present work we shall limit ourselves with one of them and shall consider model that the address of a subset $Y^{(2)} \subseteq A^{(2)}$ at the scale $\mathbf{S}^{(2)}$ is a function (in the general set theoretical meaning) defined over elements of Hilbert space that are addresses on the scale $\mathbf{S}^{(1)}$. This means, we use the approach resembling to that used for the transition from classical to quantum.

Let there the set of functions $\Omega \stackrel{def}{=} \Omega(\{\psi\})$ of quantum Hilbert space elements. Define now the set of all elements $\Omega$:

$$\Theta \stackrel{def}{=} \{\Omega\} \qquad (17)$$

Because the consideration of this Section is only an example, it allows us, also as example, to choose Θ an abstract mathematical ***space***. This choice allows us not to go away too far from the quantum mechanical formalism. What type of space? As in quantum mechanics fundamental experimental data and postulates issued from them would be necessary to answer to this question (see, for example, Dirac (1958), Shiff (1955), Landau & Lifshitz (1977)). However, at present they are absent, and so we shall consider one hypothetical way.

If Θ is chosen as a Hilbert space, the subquantum scale will be on principle like one more step following the second quantization or, which is equivalent, Fock configuration representation or Fock functionals (see, for example, Berezin (1966), Davydov (1976), Fock (1932), Fock (1937, 1934)).

Let us continue this process and transit from the scale $\mathbf{S}^{(2)}$ to the scale $\mathbf{S}^{(3)}$. By each step we suppose as before, that, the corresponding set is a ***space*** and that this space is the Hilbert one. Thus, we shall obtain the set

$$\Xi \stackrel{def}{=} \{\Theta\} \tag{18}$$

If to suppose that the set $\Xi$ is a Hilbert space, one will obtain sub-quantum scale physics as something on principle like the 4th quantization. Note that we use the term Hilbert space only for short. Really, these spaces should be supposed to be like that used in quantum mechanics, which is, generally speaking, not the Hilbert one, in particular, because of δ-functions.

## *12. NOTES ON MICROWORLD SCALES AND RELATIVISTIC QUANTUM FIELD THEORY*

This Section is dedicated to a discussion on possible applications of microworld scales' theory developed above in this work to problems of relativistic quantum field theories. We consider here how the existence and properties of microworld scales influence the relativistic quantum theory and quantum field theory. In this respect of such a consideration an important problem is whether one is allowed to "guillotiner" without heavy consequences the well-ordered set of microworld scales or such a use of the "guillotine" could lead to incorrect results. So it would be desirable, first of all, to clarify whether this sequence is in fact infinite (as it seems) or finite. We shall return below to the connection of this problem with the quantum field theory. Notice here only that it is important because such a "guillotineering" the sequence of microworld scales or *natural* limitation of the microworld scales' number (if it is finite) leads with the necessity to a certain upper limit of particle energies that is to be taken into account at summations (integrations) with respect to all possible states.

Possibly, Dirac electron – positron vacuum in his relativistic electron theory (Dirac 1958) is a kind of approximate, possible to say,

phenomenological representation of the influence of microworld scales of small hierarchy ("small" scales). A correct consideration of relativistic electron must include smaller hierarchy microworld scales and its interactions within them because the electron of very high energy penetrates into such scales (as it penetrates, for example, inside nuclei (see, for example, Hofstadter 1954, 1956, 1963, Herman & Hofstadter 1960)). In such a consideration the negative total electron energy cannot appear because the energy balance must include energies of electron interaction within all microworld scales, but not only within the *elementary particle scale (EPS)*, for which Dirac equation [6] is written. The Dirac theory can be really considered as a kind of the renormalization when the global influence of smaller than of EPS hierarchy scales is approximately taken into account by the phenomenological model of the sea of electrons with negative total energy (electron – positron vacuum). The written above *does not mean that positrons exist* in such small hierarchy scales. Something occurring within one or some such scales influences the occurring within the EPS, and in Dirac theory namely this influence is approximately expressed in terms of the electron – positron pairs' appearance, fluctuations of the electron – positron vacuum *etc.*.

Possibly, divergences, the renormalized electron charge zero (Pomeranchuk 1955, Berestetskiĭ 1976, Landau *et al* 1954) a. o. difficulties of the quantum field theory (see, for example, Bogoliubov&Shirkov 1980, Ryder 1996, Umezawa 1956, Weinberg 1995 – 2000) are originated from the influence scales "smaller" (or exactly, of hierarchy smaller) than the scale where the corresponding quantum theory of field is defined (EPS). Indeed, quantum theory of fields is a relativistic theory and so includes, as in the case of Dirac equation, very high energies of particles. Thus, summations must include (virtual) particles of such energies. Such particles are able to penetrate into scales of very small hierarchy, their behavior depends on their interaction within such scales and by this way these scales contribute to results of calculations obtained within the scale where the quantum field theory is used. Therefore, in a theory logically complete and closed into it ***all*** these scales must be taken into account. ***Probably, by this way the quantum field theory (principally very beautiful and promising one) could be "reanimated".***

*It is important for this purpose that, as it was established above, there is a scale of the minimum hierarchy (the "smallest" scale), and, by this reason, the maximum number* $q_{max}$ *of scales exists. This number "cuts" naturally the sequence of scales* **which also limits the highest value of particle energy that is to be taken into account** *at the summation with respect to all virtual states.* It could be expected that this limitation *naturally* prevents the divergences appearance.

The consideration by the quantum field theory is performed within a certain scale of the microworld (EPS). It seems we "work" all the time within this scale, all mathematical formalism is constructed just for this scale without taking into account even the existence of other scales. However, as we have just seen, the occurring within scales of smaller (than of EPS) hierarchy may influence the occurring within the EPS. How to describe this influence? It depends on the theory. If one considers models based on (elementary) particles, in some cases (but not in the general case!) one can consider transitions of such particles from the considered scale to those of smaller hierarchy and vice versa. In the framework of more general approach processes, interactions and objects that are within smaller hierarchy scales must be taken into account at consideration of their consequences detected within the scale of elementary particles.

From the proposed point of view the existing concept itself of the vacuum seems to be none other than an attempt to take into account the influence of going on within low-hierarchy scales upon considerations made within EPS by use of an approximate phenomenological model. However, the *real* vacuum (not this model!) is the set of all microworld scales.

The classification of elementary particles based on the group representation theory (see, for example, Rumer&Fet 1970) also can be affected by taking into consideration low-hierarchy microworld scales. Indeed, the symmetry can be violated by the occurring within such scales when very high energies of particles are taken into account. Roughly speaking, the interaction of particles within microworld scales of hierarchies lower than that of the EPS could be an addition to the Lagrangian breaking the symmetry existing within the EPS. This effect becomes essential when energies of particles that are to be taken into account at summations (with respect to intermediate states) are sufficiently high to let particles penetrate into such low-hierarchy ("small") scales. Strictly speaking, a correct theory must take into account **all** microworld scales. There is no matter whether, for example, the Lagrangian formalism can be used or not.

## 13. CONCLUSIONS

In the present paper the general approach is proposed to the successive penetration to well ordered sets of smaller and smaller hierarchy from the set having the highest hierarchy. This hierarchy is set up within a well ordered set of well ordered sets. Scales of the microworld are defined as a particular case of this system, and the penetration to smaller and smaller scales is replaced to the penetration to sets of lower and lower hierarchy. The substitution is necessary because in the considered case the word "small" has only intuitive meaning, if possible to say so. We simply got accustomed that atomic and nuclear scales have characteristic sizes $\sim 10^{-8}$cm and $\sim 10^{-13}$cm, correspondingly, and so we think in terms of sizes. But such an approach becomes unclear when we try to study smaller scales. The physical meaning of "small" will be lost as a consequence of the impossibility to define the concept "size" because it demands the existence of space-time, which is very questionable there. The definition of the notion "scale of the microworld" must be <u>based on physical properties</u> of occurring events. While events possess properties of a certain class, they all are within a certain scale. By this way we get rid of the use the size of a space-time region with this purpose. This approach allows one to study <u>mathematically</u> microworld scales' system as hierarchic well ordered set of well ordered sets filling these mathematical objects with the physical content.

In Secs. 2-4 we considered and developed different ways to establish hierarchy among different kinds of the information and among sets. An important way to do it is based on the consideration of the information value. An attempt is made to represent the information value mathematically, though this problem is extremely complicated because each case differs from the other ones. Three notions were defined: information value, potential information value and constrained information value. The hierarchy between different kinds of the information can be established on the grounds of the information value, or potential information value, or constrained information value, depending on the considered case.

In the case when the information is expressed in a language (natural or formal) the hierarchy can be established also on the grounds of Russell's theory of types. The formalisms developed in Secs. 2-4 serve as the mathematical ground of the microworld scales theory.

Our general mathematical approach to the problem of physical laws within different microworld scales is based on the use of two well ordered sets. Elements of one of them we have denoted *events*. We mean that in applications to the physics this term includes objects (*e. g.*, electron) and events in proper meaning occurring with them (*e. g.*, electron scattering). Elements of the second set we have denoted events' *addresses*.

Address of a subset of the set of events is a certain subset of the set of addresses put to correspondence to this subset of events by a homomorphic mapping with the feedback. The feedback is necessary to "label" the considered subset of events by the corresponding subset of the set of addresses. The mapping is realized by measurements made always by an observer **within the highest hierarchy set**. In physics it corresponds to the macroscopic observer. Indirect measurements are considered as the only type of them allowing such an observer to obtain the information on occurring within a set of low hierarchy. Limitations of our possibility to penetrate to low hierarchy sets (to small scales of the microworld, in physics) are found as issued from the multi-step character of measurements.

It was indicated that possibly quarks are not within the same scale that elementary particles, but within the following ("smaller", exactly and without inverted commas, lower hierarchy) scale where the existence of space-time continuum is questionable. If it really does not exist, coordinate transformations and their groups are nonsense within this scale, and, therefore, the theory of corresponding group representations is not fit for quarks study.

The (physical) theory is defined within each scale as a set that may be a set of operators (this option *seems* us to be the most realistic one, but for the present yet only *seems*). Then the main task is to find this set for each scale in consideration. In Sec. 9 one proposed a *hypothetical* way how to do it within different scales.

It must warn that the penetration deeper and deeper to small scales of the microworld is not a high way, but very complicated and sometimes even contradictory process. If at a certain step one finds event that cannot be understood in the framework of physical laws of the considered scale, the solution of this problem is in introduction a neighbor lower hierarchy scale with its own physical laws, <u>iff it is found out that it cannot be obtained in the framework of the considered scale by the change its physical laws.</u> So it is necessary to search for experiments and theoretical arguments to distinct between these options. Really, in the framework of each of these two options many "sub-options" exist (*e. g.*, different versions of physical laws), which may complicate essentially each step.

<u>*We want to call the attention to the fact that the existence of scales is not postulated, the proposed theory allows one find out what scales of the microworld do exist.*</u>

In the contemporary considerations scales of the hierarchy smaller than the EPS hierarchy are not considered explicitly and instead, to take nevertheless their effect into account one considers a certain model, for example, the Dirac background of electrons having the negative energy. The main difficulty of the taking into account effects produced by smaller hierarchy scales is that we do not know the going on and the physical laws within these scales. The renormalization (see, for example, Bogoliubov&Shirkov 1980, Umezawa 1956, Hepp 1969, Manukian 1983, Collins 1984, Samhofer 1999) is an attempt to exclude or to isolate this effect or, at least, its parts that could be considered as not important, being, in fact, a kind of phenomenological approach.

The similar factors, such as the taking into account interactions not only within EPS, but also within lower-hierarchy scales, could affect the classification of elementary particles on the grounds of their symmetry group representations (Rumer&Fet 1970).

Thus, relativistic quantum field theories (independently of particle energies that should be taken into account) limit the consideration with only one EPS and physical laws existing within this only scale, forgetting other scales with their physical laws. This is the main cause of well Thus, relativistic quantum field theories (independently of particle energies that should be taken into account) limit the consideration with only one EPS and physical laws existing within this only scale, forgetting other scales with their physical laws. This is the main cause of well known difficulties of these theories.

***REFERENCES***

Auletta, G., 2000, *Foundation and Interpretation of Quantum Mechanics.* World Scientific, Singapore - New Jersey - London - Hong Kong

Berestetskiĭ, V. B., 1976, Sov. Phys. Usp. **19**, 934

Berezin, F. A., 1966, *The Method of the Second Quantization*, New York, Academic Press

Bogoliubov, Nikolai N. and Shirkov, D. V., 1980, Introduction to the Theory of Quantized fields. 3rd ed., New York, Interscience,

Braginsky, V. B., & Khalili, F. Ya., 1992, *Quantum Measurement.* Cambridge University Press

Brillouin, Leon, 1956, *Science and Information Theory.* Academic Press Inc., Publishers, New York

Collins, John C., 1984, Renormalization. Cambridge: Cambridge University Press

Davydov, A. S., 1976, *Quantum Mechanics*, Oxford, Pergamon Press

Dirac, P. A. M., 1958, *The Principles of Quantum Mechanics*, Oxford at Clarendon Press

Eigen, Manfred, 1971, *Naturwissenschaften* **58**, 465-523

Fock, V., 1932, *Zeitschrift für Physik* **75**, 622-647

Fock, V., 1937, *Uchonyie Zapiski Leningradskogo Universiteta* № 17 (in Russian) (*Scientific Proceedings of Leningrad University* № 17 (1937));

Fock, V., 1934, *Sowiet Physics* **6**, 425-469

Gell-Mann, Murray&Ne'eman Yuval, 1964, *The Eightfold Way*, W. A. Benjamin, Inc., New York - Amsterdam

Gribov V. N., 2002, *Gauge Theories and Quark Confinement*, Moscow: PHASIS

Hepp, K., 1969, Théory de la renormalization. Springer-Verlag, Berlin-Heidelberg-New York

Herman, Robert and Hofstadter, Robert, 1960, High-energy electron scattering tables. Stanford, CA, Stanford University Press

Hofstadter, Robert, 1963, Electron scattering and nuclear and nucleon structure: a collection of reprints with an introduction. New York, Benjamin

Hofstadter, Robert, 1956, Rev. Mod. Phys. **28** (3), 214 – 254

Hofstadter, R., Hahn, B., Knudsen, A. W., and Mc Intyre, J. A., 1954, Phys. Rev. **95** (2), 512-515

Hosaka, Atsushi & Toki, Hiroshi, 2001, *Quarks,Baryons and Chiral Symmetry*, World Scientific Publishing Co Pte Ltd, Singapore-New Jersey-London-Hong Kong

Jech, Tomas, 2003, *Set Theory, Springer Monographs in Mathematics.* Springer-Verlag, Berlin

Kokkedee, J. J. J., *The Quark Model*, W. A. Benjamin, Inc., New York – Amsterdam, 1969

Landau, L. D., Abrikosov, A. A., Khalatnikov, I. M., 1954, Doklady Akademii Nauk SSSR (Proceedings of the Academy of Sciences of the USSR) **95,** 497 (in Russian)

Landau, L. D., & Lifshitz, E. M., 1977, *Quantum Mechanics: non-relativistic theory*, Oxford, Pergamon, Butterworth-Heinemann

Mandelstam, L. I., 1972, *Lektsii po Optike, Teorii Otnositel'nosti i Kvantovoi Mekhanike (*in Russian), Moskva, Nauka. (Lectures on Optics, Relativity Theory and Quantum Mechanics, Moscow, Nauka, 1972).

Manoukian, Edward B., 1983, Renormalization. Academic Press, New York

Nambu, Y., 1985, *QUARKS*, World Scientific Publishing Co Pte Ltd, Singapore

Ne'eman, Yuval, 1961, *Nucl. Phys.* **26**, 222 – 229

Pomeranchuk, I., 1955, Doklady Akademii Nauk SSSR (Proceedings of the Academy of Sciences of the USSR) **103**, 1005 (in Russian)

Rumer, Yu. B. & Fet A. I., 1970, Teoria Unitarnoi Simmetrii (Theory of Unitary Symmetry). "Nauka", Moscow (in Russian)

Ryder, Lewis H., 1996, Quantum Field Theory. 2$^{nd}$ ed., Cambridge: Cambridge University Press

Russell, Bertrand, 1908, *American J. Math.* **30**, 222-262

Samhofer, Manfred, 1999, Renormalization. Berlin, Springer

Shannon, C. E., 1948, *The Bell System Technical Journal* **27**, 379-423, 623-656

Shiff, L. I., 1955, *Quantum Mechanics*, McGrow-Hill Book Company, Inc., New York-Toronto-London

Umezawa, H., 1956, Quantum Field Theory. North-Holland Publishing Company, Amsterdam

Volkenstein, M. V., 1977, *Found. Phys.* **7**, 97-109

Weinberg, Steven, 1995 – 2000, The Quantum Theory Of Fields. Cambridge: Cambridge University Press, vol. I - iii

Whithead, Alfred North & Russell, Bertrand, 1963, *Principia Mathematica*, Second Edition, Cambridge at the University Press